\pdfoutput=1

\documentclass[%
 reprint,
 showpacs,
 nofootinbib,
 amsmath,amssymb,
 aps,
 pra,
]{revtex4-1}

\usepackage{graphics}
\usepackage{graphicx}
\usepackage{color}
\usepackage{amsmath}
\usepackage{slashed}
\usepackage{epsfig}
\usepackage{hyperref}

\begin{document}

\title{Electroweak naturalness in three-flavour Type I see-saw\\and implications for leptogenesis}

\author{Jackson D. Clarke, Robert Foot, Raymond R. Volkas}
\affiliation{ARC Centre of Excellence for Particle Physics at the Terascale, \\ 
School of Physics, University of Melbourne, 3010, Australia.}

\date{\today}

\begin{abstract}

In the Type I see-saw model, 
the naturalness requirement that corrections to the electroweak $\mu$ parameter not exceed 1~TeV
results in a rough bound on the 
lightest right-handed neutrino mass, $M_{N_1}\lesssim 3\times 10^7$~GeV.
In this letter we derive generic bounds applicable in \textit{any} three-flavour Type I see-saw model.
We find $M_{N_1}\lesssim 4\times 10^7$~GeV and $M_{N_2}\lesssim 7\times 10^7$~GeV.
In the limit of one massless neutrino, there is no naturalness bound 
on $M_{N_3}$ in the Poincar\'e protected decoupling limit.
Our results confirm that no Type I see-saw model can
explain the observed neutrino masses and baryogenesis via hierarchical
($N_1$-, $N_2$-, or $N_3$-dominated) thermal leptogenesis while remaining completely natural.

\end{abstract}


\maketitle

\section{Introduction}

The discovery of a particle consistent with the 
standard model (SM) Higgs Boson \cite{Aad2012tfa,*Chatrchyan2012ufa} 
appears to confirm the standard mechanism of electroweak spontaneous symmetry breaking:
The Higgs field $\phi$ gains a vacuum expectation value $\langle\phi\rangle$
by virtue of the potential $V=-\mu^2|\phi|^2+\lambda|\phi|^4$
at a scale set by the renormalised parameter $\mu = m_h/\sqrt{2} \approx 88$~GeV.
Much of modern high energy physics has been concerned with the naturalness of this scale.
The fact remains that the SM alone (without gravity) suffers no hierarchy problem;
a large cancellation between an unmeasurable bare parameter 
and an unphysical cutoff scale can be assigned no physical significance.
Indeed, if no physical large scale exists there can be no hierarchy problem.
Neither is the presence of such a scale sufficient 
for a hierarchy problem \cite{Bardeen1995kv,Foot2013hna}.\footnote{See Ref.~\cite{Foot2013hna} 
for citations to other literature on this topic.}
To find out, the pragmatic physicist should just take
a model, explicitly calculate corrections,
and express them in terms of (in principle) measurable parameters.
If those corrections are large compared to measured values,
only then could naturalness become a concern.

Vissani did just that in the one-flavour Type I see-saw model \cite{Vissani1997ys}
(see also Refs.~\cite{Casas2004gh,*Abada2007ux,*Xing2009in,*Bazzocchi2012de,*Farina2013mla,*deGouvea2014xba,*Davoudiasl2014pya}).
He calculated a correction to the electroweak $\mu$ parameter
\begin{align}
 \delta\mu^2 \approx \frac{1}{4\pi^2}\frac{1}{\langle\phi\rangle^2}
   m_\nu M_N^3 , 
\end{align}
where $\langle\phi\rangle\approx174$~GeV.
If required to be less than 1~TeV$^2$, a neutrino of mass $m_{atm}\approx 0.05$~eV
implies an upper bound on the right-handed neutrino mass, $M_N\lesssim 3\times 10^7$~GeV.
Nevertheless, the model still provides an elegant explanation
of the smallness of the neutrino mass scale \cite{Minkowski1977sc,*Mohapatra1979ia,*Yanagida1979as,*GellMann1980vs}. 
The Type I see-saw model is also capable of explaining
baryogenesis via leptogenesis,
the Fukugita-Yanagida mechanism \cite{Fukugita1986hr}.
The standard version requires the lightest right-handed neutrino
to satisfy $M_{N_1}\gtrsim 5\times 10^8$~GeV \cite{Davidson2002qv,Giudice2003jh},
in obvious tension with the naturalness bound above.

Thus it appears that one cannot use the Type I see-saw model to explain both
the observed neutrino masses and baryogenesis via standard thermal leptogenesis
without ceding naturalness.
This letter aims to establish whether this conclusion holds in the 
three-flavour Type I see-saw model in full generality.

\begin{figure}[t]
 \centering
 \includegraphics[width=0.65\columnwidth]{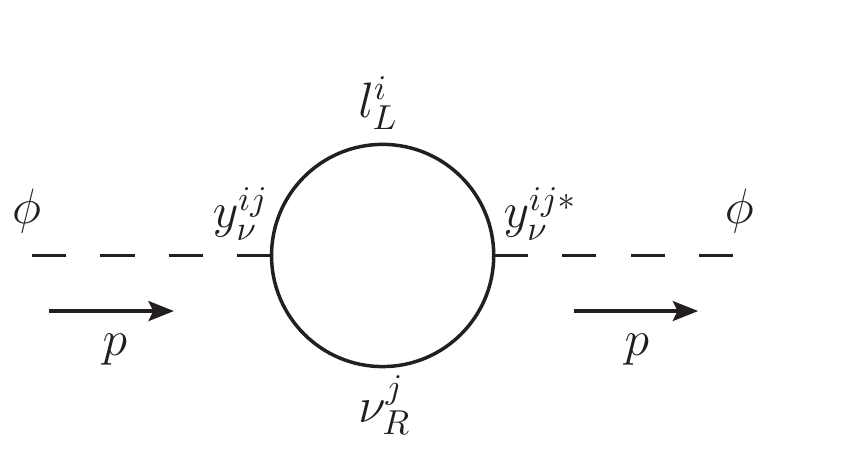}
 \caption{Loop diagram leading to $\delta\mu^2$.}
 \label{figmucorrx}
\end{figure}

\section{Preliminaries}

The pertinent part of the Type I see-saw Lagrangian is
\begin{align}
 -\mathcal{L} = \overline{l_L^i} y_e^{ij} e_R^j \phi + \overline{l_L^i} y_\nu^{ij} \nu_R^j \tilde{\phi} + \frac12 \overline{(\nu_R^i)^c} M_N^{ij} \nu_R^j + h.c. \;,
\end{align}
where $i,j$ are flavour indices, $l_L^i \equiv (\nu_L^i, e^i_L)^T$, and $\tilde\phi \equiv i\tau_2\phi^*$.
One is free to rotate and rename the fields such that
$M_N \equiv \mathcal{D}_{M} = \text{diag}(M_1,M_2,M_3)$ has real positive diagonal entries.

After symmetry breaking, and if $M_j\gg y_\nu^{ij}\langle\phi\rangle$, the Lagrangian becomes
\begin{align}
 -\mathcal{L} \approx \overline{l_L^i} y_e^{ij} e_R^j \langle\phi\rangle + 
   \frac12 \overline{\nu_L^i} m_\nu^{ij} (\nu_L^j)^c + 
   \frac12\overline{(\nu_R^i)^c} M_i \nu_R^i + h.c. \;,
\end{align}
where $m_\nu = \langle\phi\rangle^2 y_\nu \mathcal{D}_M^{-1} y_\nu^T$ 
is the neutrino mass matrix.
One can diagonalise $m_\nu$ with a unitary matrix $U$,
\begin{align}
 \mathcal{D}_{m} \equiv \text{diag}(m_1,m_2,m_3) = U m_\nu U^T,
\end{align}
where $m_i$ are the neutrino masses.
Following Casas-Ibarra \cite{Casas2001sr}, it is possible to express $y_\nu$ as
\begin{align}
 y_\nu = \frac{1}{\langle\phi\rangle}
   U^\dagger \mathcal{D}_m^\frac12 R \mathcal{D}_M^\frac12, \label{eqCasasIbarra}
\end{align}
where $R$ is a (possibly complex) orthogonal ($R^T R = R R^T = \mathbb{I}$) matrix.
$R$ is physically relevant and measurable in principle
(e.g. by studying the production and decays of the $\nu_R^j$),
however measurements to date tell us nothing about it.


\begin{figure}[t]
 \centering
 \includegraphics[width=0.95\columnwidth]{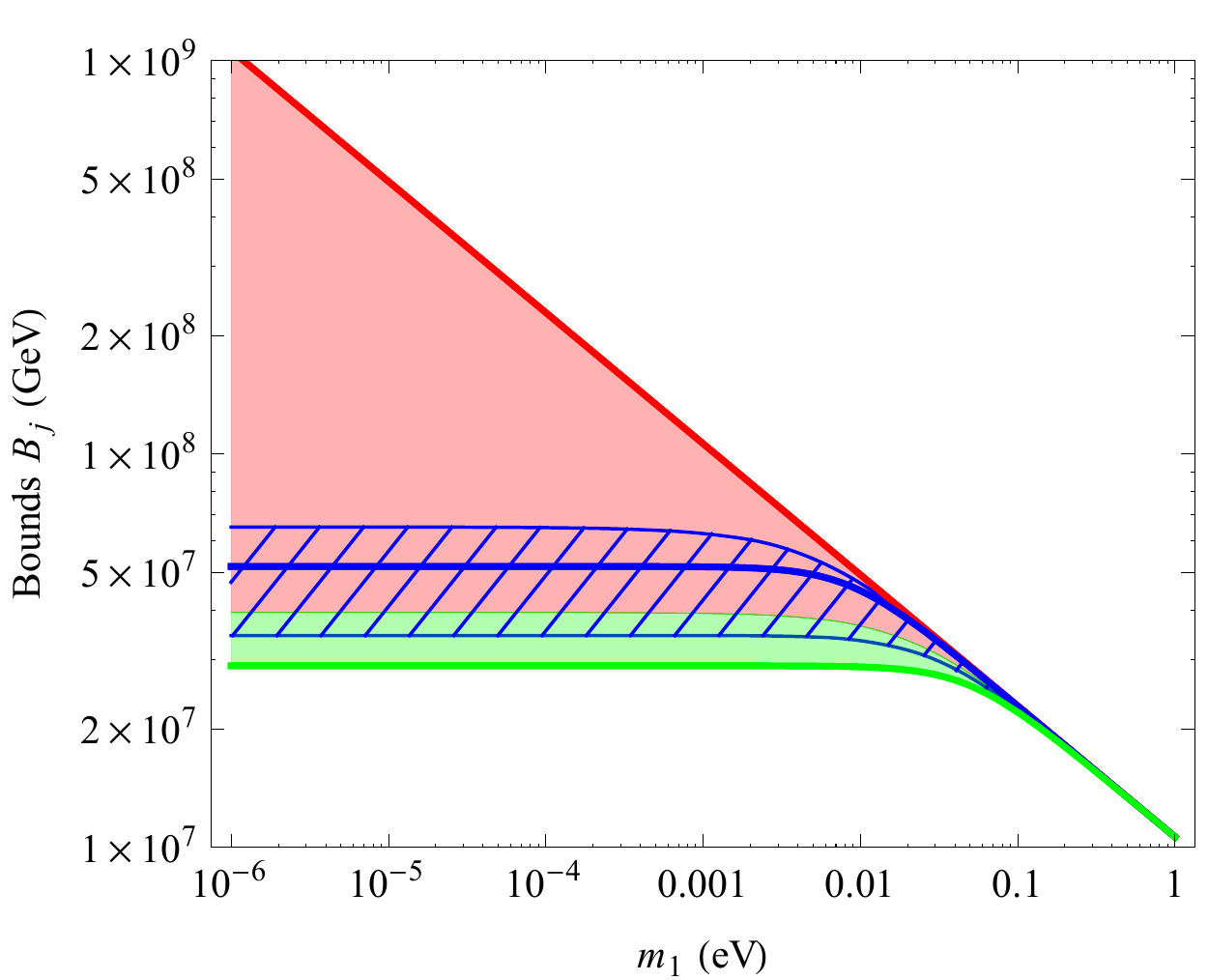}
 \caption{As a function of the lightest neutrino mass in NO,
 shown as darker/hatched/lighter (red/blue-hatched/green colour online) 
 is the region of attainable values for the $B_3\ge B_2\ge B_1$ 
 upper bounds on right-handed neutrino masses, by requiring the corrections to the electroweak $\mu$ parameter 
 be no greater than 1~TeV (Eq.~\ref{eqdmult1TeV}). 
 The regions assume the orthogonal matrix $R$ is real.
 Thick solid lines show the case when $R=\mathbb{I}$.
 The case for complex $R$ is similar, except there is no lower limit to the $B_j$ (see text).
 }
 \label{figNObounds}
\end{figure}

From Fig.~\ref{figmucorrx}, we calculate the correction to $\mu^2$ in $\overline{MS}$ scheme as
\begin{align}
 \left| \delta \mu^2 \right| = \frac{1}{4\pi^2} y_\nu^{ij} M_j^2 
   \left| \log\left[\frac{M_j}{\mu_R}\right]  - \frac14 \right|
   y_\nu^{ij*} + \mathcal{O}\left(\mu^2\right),
\end{align}
where $\mu_R$ is the renormalisation scale.
The renormalisation group equation for $\mu^2(\mu_R)$ will receive
a contribution $\approx \frac{1}{4\pi^2}|y_\nu^{ij}|^2M_j^2$.
If this contribution is much larger than the electroweak scale,
only a very finely tuned $\mu^2(\mu_R\gg M_j)$ will achieve $m_h\sim 125$~GeV;
the natural scale for $m_h$ is $\sim|y_\nu^{ij}| M_j$.

Taking the quantity within absolute values to be unity,
the contribution from all nine diagrams becomes
\begin{align}
 \left| \delta \mu^2 \right| \approx \frac{1}{4\pi^2} \text{Tr}\left[ y_\nu \mathcal{D}_M^2 y_\nu^\dagger \right] .
\end{align}
Upon substitution of the Casas-Ibarra form (Eq.~\ref{eqCasasIbarra}), one obtains the simple relation
\begin{align}
 \left| \delta \mu^2 \right| \approx \frac{1}{4\pi^2}\frac{1}{\langle\phi\rangle^2} 
   \text{Tr}\left[ \mathcal{D}_m R \mathcal{D}_M^3 R^\dagger \right] . \label{eqdmu}
\end{align}
Note that there is no explicit dependence on $U$, as one could anticipate,
since all of $U$ can be absorbed by $l_L\to U l_L$, $y_\nu \to U y_\nu$.
One ends up with three positive-definite corrections proportional to the cube of each heavy neutrino mass.
Naturalness demands that these corrections each be less than some scale not far above $\mu\approx88$~GeV.
In our calculations we therefore require the three bounds: 
\begin{align}
 \frac{1}{4\pi^2}\frac{1}{\langle\phi\rangle^2} M_j^3 \sum_i m_i |R_{ij}|^2 < 1\text{ TeV}^2 , \nonumber \\
 \Rightarrow M_j \lesssim 2.9\times 10^7 \text{ GeV}
   \left( \frac{0.05\text{ eV}}{\sum_i m_i |R_{ij}|^2} \right)^\frac13 , \label{eqdmult1TeV}
\end{align}
where $R_{ij}$ are the entries of $R$.
Our results can be easily rescaled for a different naturalness criterion.

\begin{figure}[t]
 \centering
 \includegraphics[width=0.95\columnwidth]{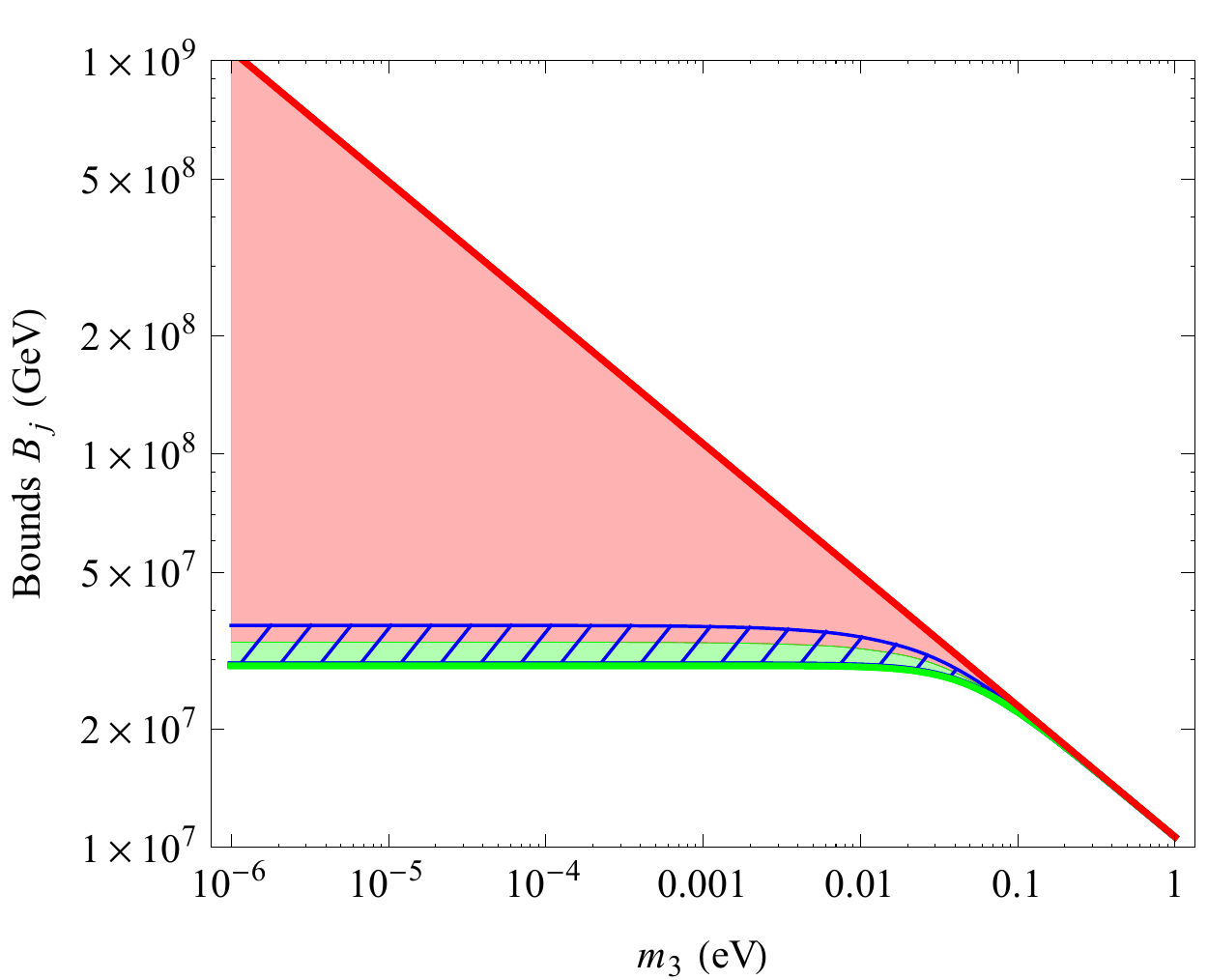}
 \caption{As in Fig.~\ref{figNObounds} but as a function of the lightest neutrino mass in IO. 
 Note that the thick blue line is obscured by the thick green line.}
 \label{figIObounds}
\end{figure}

\section{Results}

Eq.~\ref{eqdmult1TeV} results in three upper bounds on the right-handed neutrino masses.
It says nothing about their mass ordering, since one can always append to $R$ 
a permutation matrix.
However we can always order the bounds by their size;
we will call them $B_j$ and take $B_1\le B_2\le B_3$.

We are interested in the values of $B_j$ attainable from Eq.~\ref{eqdmult1TeV}.
Thus all we have to do is extremise these bounds over $R$.
We used the mass squared differences of \textsc{NuFIT v2.0} \cite{GonzalezGarcia2014bfa},
\begin{align}
\Delta m_{21}^2 = 7.50\times10^{-5}\text{ eV}^2, \nonumber \\ 
\Delta m_{3l}^2=\pm 2.46\times 10^{-3}\text{ eV}^2,
\end{align}
where $\Delta m​_{3l}^2=\Delta m​_{31}^2> 0$ for normal ordering (NO) 
and $\Delta m​_{3l}^2=\Delta m​_{32}^2 < 0$ for inverted ordering (IO), 
and treat
the lightest neutrino mass ($m_1$ for NO or $m_3$ for IO) as unknown.
The $B_j$ were numerically extremised  over a 
parameterisation of $R$.
The results were checked analytically and with scatterplots.
Figs.~\ref{figNObounds} and \ref{figIObounds} show the case 
for NO and IO when $R$ is real.
The solid lines are for $R=\mathbb{I}$.

The first thing to notice is that as the lightest neutrino mass
tends to zero, the largest bound $B_3$ can potentially evaporate.
This only happens in models where $R$ is of a particular form, e.g. in NO,
as is evident from Eq.~\ref{eqdmult1TeV},
\begin{align}
 R = \left(
 \begin{array}{ccc}
  R_{11} & R_{12} & \pm 1 \\
  R_{21} & R_{22} & 0 \\
  R_{31} & R_{32} & 0
 \end{array}
 \right) \label{eqB3large}
\end{align}
or some column permutation, where $R_{11}=R_{12}=0$ if $R$ is real.
This corresponds to the Poincar\'e protected decoupling limit $y_\nu^{i3}\to 0$
and an effective two-flavour see-saw \cite{Frampton2002qc,*Raidal2002xf,*Guo2006qa}.

The maximisation of $B_2$ occurs when $B_2=B_3$ and corresponds in NO to $R$ of the form
\begin{align}
 R = \pm \left(
 \begin{array}{ccc}
  0 & \frac{1}{\sqrt{2}} & -\frac{1}{\sqrt{2}} \\
  0 & \frac{1}{\sqrt{2}} & \frac{1}{\sqrt{2}} \\
  \pm 1 & 0 & 0
 \end{array}
 \right),
\end{align}
up to column permutations. Similarly the minimisation of $B_2$ occurs when $B_1=B_2$, and corresponds to
\begin{align}
 R = \pm\left(
 \begin{array}{ccc}
  0 & 0 & \pm1 \\
  \frac{1}{\sqrt{2}} & -\frac{1}{\sqrt{2}} & 0 \\
  \frac{1}{\sqrt{2}} & \frac{1}{\sqrt{2}} & 0
 \end{array}
 \right). 
\end{align}
The maximisation (minimisation) of $B_1$ ($B_3$) occurs when $B_1=B_2=B_3$.
This corresponds to a conspiratorial form of $R$.

Even though these arrangements are possible, it is clear
from Figs.~\ref{figNObounds} and \ref{figIObounds} that it is not possible
to construct a Type I see-saw model that changes the bounds $B_1$ and $B_2$ by 
more than a factor of 2 when $R$ is real.
Even if one does saturate these bounds, it is not possible then
to place the right-handed neutrino masses at this bound and 
maintain a hierarchy that is the basis of many of the calculations
for thermal leptogenesis.

In the case of $R$ complex the upper limits of the $B_j$ 
are the same as the $R$ real case.
However the lower limits can potentially be much lower.
The reason is that complex $R$ with entries of arbitrarily large magnitude exist.
Let us illustrate this in the two-flavour case.
An example is
\begin{align}
 R = \left(
 \begin{array}{cc}
  \cosh x   & i\sinh x \\
  -i\sinh x & \cosh x
 \end{array}
 \right).
\end{align}
In this case,
\begin{align}
 Uy_\nu = \frac{1}{\langle\phi\rangle}
   \left(
   \begin{array}{cc}
    \sqrt{m_1 M_1} \cosh x & - i \sqrt{m_1 M_2} \sinh x \\
    i \sqrt{m_2 M_1} \sinh x & \sqrt{m_2 M_2} \cosh x
   \end{array}
   \right) .
\end{align}
If $\cosh x \gg 1$, one need only calculate $m_\nu$ to see that
the smallness of neutrino masses is only explained by
fortuitous cancellations between entries of $y_\nu$ 
that constitute a fine tuning.
If we demand that the entries of $R$ have magnitude not exceeding 1,
then the results for complex $R$ are essentially the same as in the real case.
In general, however, allowing complex $R$ can only degrade the attainable region for the $B_j$.

\section{Conclusion}

Corrections to the $\mu$ parameter in the 
Type I see-saw model can be expressed in the concise form of Eq.~\ref{eqdmu},
as a function of an unknown (but in principle measurable) orthogonal matrix $R$.
Requiring these corrections to be less than 1~TeV results in
three bounds on the right-handed neutrino masses, $B_1 \le B_2 \le B_3$.
As shown in Figs.~\ref{figNObounds} and \ref{figIObounds}, we 
find that $B_1$ and $B_2$ can be varied by no more than a factor 2
around their values in the $R=\mathbb{I}$ case.
The bound $B_3$ evaporates in models with a massless neutrino
and $R$ of the form Eq.~\ref{eqB3large} (up to column permutations) for NO, 
and similarly for IO. 
This corresponds to the Poincar\'e protected decoupling limit $y_\nu^{i3}\to 0$.
In short, we obtain the generic bounds
\begin{subequations}
\begin{align}
 M_{N_1} &\lesssim 4\times 10^7 \text{ GeV}, \label{eqboundN1} \\
 M_{N_2} &\lesssim 7\times 10^7 \text{ GeV}, \label{eqboundN2} \\
 M_{N_3} &\lesssim 3\times 10^7 \text{ GeV}\left(\frac{0.05\text{ eV}}{m_{min}}\right)^\frac13, \label{eqboundN3} 
\end{align}
\end{subequations}
where $m_{min}$ is the lightest neutrino mass.
For a given model, however, the bounds will be more stringent.

Baryogenesis via standard ($N_1$-dominated, hierarchical)
thermal leptogenesis requires $M_{N_1}\gtrsim 5\times10^8$~$(2\times 10^9)$~GeV 
for $N_1$ with thermal (zero) initial abundancy \cite{Giudice2003jh}\footnote{These
bounds are unaffected by flavour considerations \cite{Blanchet2006be,*JosseMichaux2007zj}.},
in conflict with Eq.~\ref{eqboundN1}.

In $N_2$ leptogenesis, it is possible to have $M_{N_1}\lesssim 10^7$~GeV.
There are two scenarios.
One is in the $N_1$-decoupling limit \cite{DiBari2005st},
and the other relies on special flavour alignments to protect an $N_2$-generated
asymmetry from $N_1$ washout \cite{Vives2005ra,*Engelhard2006yg}.
Both are in conflict with Eq.~\ref{eqboundN2},
as such a light $N_2$ is unable to produce the required
asymmetry for the usual reasons \cite{Davidson2002qv,Giudice2003jh}.


One might think that there is still room left for $N_3$ leptogenesis.
This turns out to not be the case.
In order to naturally have $M_{N_3}\gtrsim 10^9$~GeV, one must have $m_{min}\lesssim 10^{-6}$~eV and
$R$ in a decoupling limit such as Eq.~\ref{eqB3large}.
However in this limit the CP asymmetry from $N_3$ decays is \cite{DiBari2005st}
\begin{align}
 \varepsilon_3 \sim 10^{-1} \sum_{i=1,2} \frac{m_{min}M_i^2}{\langle\phi\rangle^2 M_3} \text{Im}\left(R_{1i}^2\right),
\end{align}
which is far too small.

Thus our results confirm that no minimal Type I see-saw model can explain the neutrino masses
and baryogenesis via hierarchical ($N_1$-, $N_2$-, or $N_3$-dominated) thermal leptogenesis 
while remaining completely natural.
In the minimal scenario, the only ways to avoid this conclusion
are to assume $N_1$ has dominant initial abundancy 
(in this case leptogenesis is possible with $M_{N_1}\gtrsim 2\times 10^7$~GeV \cite{Giudice2003jh},
which is marginally consistent with Eq.~\ref{eqboundN1}),
allow a resonant enhancement \cite{Pilaftsis2003gt}, or consider an entirely different mechanism \cite{Akhmedov1998qx}.

\acknowledgments

This work was supported in part by the Australian Research Council.

\bibliography{references}

\end{document}